# Tailoring the light-matter interaction for high-fidelity holonomic gate operations in multiple systems

ZHIHUANG KANG[1,2,3], SHUTONG WU[1,2,3], KUNJI HAN[1,2,3], JIAMIN QIU,[1,2,3], JOEL MOSER[1,2,3], JIE LU[4,5] AND YING YAN,[1,2,3,*]

*1 School of Optoelectronic Science and Engineering & Collaborative Innovation Center of Suzhou Nano Science and Technology, Soochow University, 215006 Suzhou, China.*
*2 Key Lab of Advanced Optical Manufacturing Technologies of Jiangsu Province & Key Lab of Modern Optical Technologies of Education Ministry of China, Soochow University, 215006 Suzhou, China.*
*3 Engineering Research Center of Digital Imaging and Display, Ministry of Education, Soochow University, 215006 Suzhou, China.*
*4 Department of Physics, Shanghai University, 200444 Shanghai, China.*
*5 Shanghai Key Lab for Astrophysics, 100 Guilin Road, 200234 Shanghai, China.*
*\*yingyan@suda.edu.cn*

**Abstract:** Realization of quantum computing requires the development of high-fidelity quantum gates that are resilient to decoherence, control errors, and environmental noise. While non-adiabatic holonomic quantum computation (NHQC) offers a promising approach, it often necessitates system-specific adjustments. This work presents a versatile scheme for implementing NHQC gates across multiple qubit systems by optimizing multiple degrees of freedom using a genetic algorithm. The scheme is applied to three qubit systems: ensemble rare-earth ion (REI) qubits, single REI qubits, and superconducting transmon qubits. Numerical simulations demonstrate that the optimized gate operations are robust against frequency detuning and induce low off-resonant excitations, making the scheme effective for advancing fault-tolerant quantum computation across various platforms.

## 1. Introduction

Benefiting from the properties of quantum superposition and quantum entanglement, quantum computation is emerging as a revolutionary model for addressing complex problems that are challenging for classical computation, such as fast and efficient prime factorization, simulating the behavior of molecules at a quantum level, and unsorted database searching [1-3]. A critical step towards realizing the immense computational capability of quantum computation is developing high-fidelity quantum gates to perform quantum algorithms within the quantum system used for the computation [4]. The process heavily relies on the precise control and tailoring of the light-matter interaction, which inevitably faces challenges such as decoherence, control errors, and environmental noises during the experimental period [5,6].

To achieve more accurate and robust quantum computation, geometric quantum computation based on the geometric phase has been proposed [7]. In this approach the dynamical phase is either removed, canceled, or made proportional to the geometric phase [8,9]. Geometric phases, dependent on the global properties of the evolution path, inherently offer tolerance to some control errors [10]. However, geometric quantum computation based on adiabatic evolution is slow due to the limitation of the adiabatic condition, leading to severe degradation in gate operational fidelity as a result of decoherence from the interaction between the quantum system and the environment. To speed up the evolution, non-adiabatic geometric quantum computing (NGQC) [11,12] and non-adiabatic holonomic quantum computation (NHQC) were proposed [13,14], and soon experimentally implemented in various quantum systems [15-18]. However, the original scheme for generating an arbitrary single-qubit gate in a three-level $\Lambda$ configuration requires two loops of state evolution [13]. To reduce the complexity of utilizing two loops, the single-shot scheme [19,20], and the single-loop multiple-pulse scheme in a resonant model were developed [21]. The advantage of both schemes is that an arbitrary single-qubit gate can be realized in a single loop while maintaining full flexibility in the choice of the pulse shape and pulse duration. More information on NHQC can be found in recent review articles [22,23].

Among the various schemes proposed for quantum computation, significant progress has been made in quantum gate optimization, with advanced techniques developed to mitigate the impact of noise and control errors. Universal quantum optimal control techniques, including machine-learning-assisted control [24], closed-loop optimization [25], and error-robust control methods [26], have been proposed to improve the quantum operation fidelity in noisy environments. Machine-learning algorithms optimize quantum gates by learning the system dynamics and adjusting control parameters in real-time to mitigate random noise, offering a solution to real-world imperfections that are difficult to model analytically. Closed-loop optimization refines control parameters iteratively during experiments, ensuring high-fidelity quantum operations. Additionally, error-robust control strategies improve the resilience of quantum gates, which is crucial for large-scale quantum computing.

Achieving high-fidelity gate operations in a specific experimental system is demanding, as the light-matter interaction is not only exposed to decoherence, control errors, and environmental noises, but also to the physical limitations or constraints inherent to that system. To illustrate these challenges, we consider three experimental systems: an ensemble rare-earth ions (REI) system, a single REI system, and a superconducting transmon qubits system. The REI system serves as a competitive test bed for quantum computing and quantum memories, owing to its exceptional optical and spin coherence properties. The coherence time of the qubits can reach up to 6 hours [27], while the optical coherence time can be several milliseconds [28]. In an ensemble rare-earth ions (REI) system, such as a randomly doped $Pr^{3+}$: $Y_2SiO_5$ crystal, a qubit is represented by a group of $Pr^{3+}$ ions that reside in a clean spectral pit. Their optical transition frequencies are inhomogeneously spread over an interval of 170 kHz [29], with the center of the transition about 3.5 MHz away from the edge of the pit, where many other ions are present [30]. These characteristics impose two constraints on the gate operations: (i) the light-matter interaction should be robust against the frequency detuning within a 170 kHz interval so that the qubit ions act as one, and (ii) it should not affect the state of the ions that sit at the edge of the spectral pit, inducing negligible off-resonant excitation on those ions. Y. Yan *et al.* [30,31] applied the Lewis-Riesenfeld invariant theory to the REI system and successfully suppressed the infidelity of creating an arbitrary superposition state by a factor of five through optimizing the pulse shape. In a single REI system, the robustness requirement is not present any longer, but the low off-resonant excitation constraint remains. For instance, single $Eu^{3+}$ ions in a $Eu^{3+}$: $Y_2SiO_5$ crystal [32,33], the off-resonant excitation occurs at frequency detuning larger than 8.9 MHz due to the level structure of $Eu^{3+}$ ions. Adam *et al.* developed an arbitrary single-qubit gate operation by utilizing two two-color Gaussian pulses with controllable phases to reduce the state leakage and phase errors caused by off-resonant excitations [33].

The superconducting transmon qubits system has emerged as a leading candidate for fault tolerant quantum computing, offering relatively long coherence times and excellent scalability [34]. In the context of surface code quantum computing employing superconducting transmon qubits, the robustness condition is critical. Since implementing a fault-tolerant logical qubit can require thousands of physical qubits [35], initialization and control of these physical qubits in a superconducting circuit is complex due to slight variations in their addressing frequencies resulting from fabrication limitations [36]. Therefore, the light-matter interaction in this scenario must be robust against variations in addressing frequency. While various schemes meet the specific requirements in each system, it is more efficient and crucial to develop a unified gate operation scheme for advancing high-fidelity quantum gate operations across diverse qubit systems. This may provide significant support for the platforms that integrate multiple quantum systems and promote the development of qubit networks.

Here, we propose a theoretical scheme for constructing holonomic gates applicable in multiple systems by simply adjusting the multiple degrees of freedom available in the model, based on the single-loop multiple-pulses NHQC approach [21]. We applied the scheme to three different qubit systems: an ensemble REI system, a single REI system, and a superconducting qubit system, and performed numerical simulations to verify the performance of the quantum gates under physical constraints. Numerical simulations based on Lindblad master equation [37] showed that high-fidelity gate operations are achieved in each system despite the respective physical constraints. Therefore, our scheme provides a more convenient and efficient approach to addressing the specific limitations of each system, ultimately enabling high-fidelity control across various physical systems.

The organization of the article is as follows. In Section 2, we introduce the theoretical model for implementing universal arbitrary holonomic gates, the ansatz of the pulse, and the optimization method. Section 3 presents the numerical simulation results of the gate operations' performance across three quantum systems. Section 4 offers a discussion and conclusion of the work. Finally, the appendix provides details on the optimization process, information about compensation pulses, and more performance matrices of other gates.

## 2. Theoretical model

The resonant three-level system under consideration is depicted in Fig. 1(a), in which two Rabi frequencies, $\Omega_0(t)$ and $\Omega_1(t)$, are utilized to couple the qubit levels $|0\rangle$ and $|1\rangle$ to the excited state $|e\rangle$, respectively. $\Omega_{0,1}(t)$ is defined as $\Omega_{0,1}(t) = -\vec{\mu} \cdot \vec{E}_{0,1}/\hbar$, where $\vec{\mu}$ represents the atomic transition dipole moment and $\vec{E}$ the external electric field. In this work, we let $\Omega_0(t) = 2\sin(\theta/2)\Omega(t)e^{i\varphi_0}$ and $\Omega_1(t) = -2\cos(\theta/2)\Omega(t)e^{i\varphi_1}$, with $\theta$ and $\varphi_{0,1}$ being time-independent angles in the range of $[0, 2\pi]$. $\Omega(t)$ represents the shared component of the time-dependent envelope of the Rabi frequencies. Under the rotating-wave approximation [38], the interaction Hamiltonian of the three-level system can be expressed in the basis of $\{|0\rangle, |e\rangle, |1\rangle\}$ as follows (assume $\hbar = 1$ hereafter)

$$H(t) = \frac{1}{2}\begin{bmatrix} 0 & \Omega_0(t) & 0 \\ \Omega_0^*(t) & 0 & \Omega_1^*(t) \\ 0 & \Omega_1(t) & 0 \end{bmatrix}, \tag{1}$$

where $\Omega^*_{0,1}(t)$ indicates the complex conjugate of $\Omega_{0,1}(t)$.

If one rotates the base of the three-level system from $\{|0\rangle, |e\rangle, |1\rangle\}$ to $\{|b\rangle, |e\rangle, |d\rangle\}$ as follows

$$\begin{bmatrix} |b\rangle \\ |e\rangle \\ |d\rangle \end{bmatrix} = \begin{bmatrix} \sin\frac{\theta}{2}e^{i\varphi_0} & 0 & -\cos\frac{\theta}{2}e^{i\varphi_1} \\ 0 & 1 & 0 \\ \cos\frac{\theta}{2}e^{-i\varphi_1} & 0 & \sin\frac{\theta}{2}e^{-i\varphi_0} \end{bmatrix} \cdot \begin{bmatrix} |0\rangle \\ |e\rangle \\ |1\rangle \end{bmatrix}, \tag{2}$$

the Hamiltonian of the system will reduce to

$$H(t) = \Omega(t)(|b\rangle\langle e| + |e\rangle\langle b|). \tag{3}$$

One can see that $|d\rangle$ is currently decoupled from the external light field, called the dark state. Its orthogonal state $|b\rangle$ is named as bright state. Therefore, the original three-level system is now turned to an equivalent two-level system, as shown in Fig.1(b). In the subsequent passage, we will elucidate the construction of the gate operations in this new base.

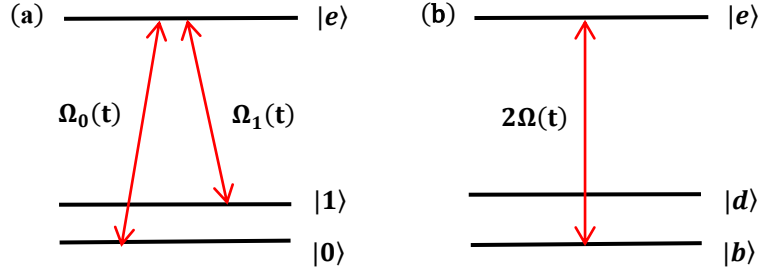

Fig. 1. Schematic energy level of a three-level system (a) in the basis of $\{|0\rangle, |e\rangle, |1\rangle\}$ and (b) in the basis of $\{|b\rangle, |e\rangle, |d\rangle\}$. The qubit is represented by two ground states $|0\rangle$ and $|1\rangle$, which are coupled to excited state $|e\rangle$ through optical transitions $|0\rangle - |e\rangle$ and $|1\rangle - |e\rangle$ with Rabi frequency of $\Omega_0(t)$ and $\Omega_1(t)$, respectively. In (b) the original three-level system reduces to an equivalent two-level system, where dark state $|d\rangle$ is decoupled from the external light field.

## 2.1 Arbitrary Holonomic Gates

Here, we demonstrate how to construct the set of universal arbitrary holonomic gates using the single-loop multiple-pulses scheme. In theory, the single-loop could be divided to $L$ segments ($L \geq 2$). Each segment involves one pulse pair which has a definite pulse area and phase relationship. However, the implementation in practice becomes increasingly complex as $L$ increases. As an illustration, here we focus on the case $L=2$, where $\Omega(t)$, $\varphi_0$ and $\varphi_1$ in each segment are as follows

$$\int_0^\tau \Omega(t)\, dt = \frac{\pi}{2}, \varphi_0 = -\phi, \varphi_1 = 0, t \in [0, \tau] \tag{4}$$

$$\int_\tau^{2\tau} \Omega(t)\, dt = \frac{\pi}{2}, \varphi_0 = -\phi + \beta + \pi, \varphi_1 = \beta + \pi, t \in [\tau, 2\tau], \tag{5}$$

where $\tau$ denotes the duration of each pulse-pair, $\phi$ the relative phase between the two pulses within a pulse-pair, and $\beta$ the phase difference between the two segments. The two Rabi frequencies $\Omega_0(t)$ and $\Omega_1(t)$ drive the quantum state evolving from $|b\rangle$ to $|e\rangle$ along path 1, and back to $|b\rangle$ along path 2, as illustrated in Fig. 2. The phase difference between the first and second segments is $\beta$, which is exactly half of the solid angle enclosed by the loop. During the course of evolution, the dynamical phase at any given time, $\gamma_d(t) = -\int_0^t \langle k(t')|H(t')|l(t')\rangle dt'$, is zero as $\langle k(t')|H(t')|l(t')\rangle \equiv 0$ for $k, l = b, d$. This implies that the gate is purely of the geometric nature.

The time evolution operator of the system is $U(\tau, 0) = \hat{P} e^{-i\int_0^\tau H(t)dt} = e^{-i\int_0^\tau H(t)dt}$, where the time ordering operator $\hat{P}$ is diminished since the Hamiltonian in Eq. (1) at any two distinct times commute with each other. Under the constraints shown in Eqs. (4) and (5), the bright and dark states evolve as follows:

$$|b(2\tau)\rangle = U(2\tau,\tau)U(\tau,0)|b\rangle = e^{i\beta}|b\rangle, \tag{6}$$
$$|d(2\tau)\rangle = U(2\tau,\tau)U(\tau,0)|d\rangle = |d\rangle. \tag{7}$$

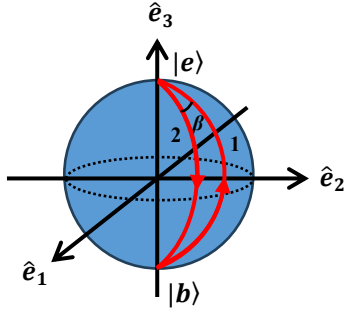

Fig. 2. Evolution path of the single-loop two-pulses NHQC approach on the Bloch sphere.

Therefore, the time evolution operator in the qubit space reads

$$U(2\tau,0) = e^{i\frac{\beta}{2}}\begin{bmatrix} \cos\frac{\beta}{2} - i\sin\frac{\beta}{2}\cos\theta & -i\sin\frac{\beta}{2}\sin\theta e^{-i\varphi} \\ -i\sin\frac{\beta}{2}\sin\theta e^{i\varphi} & \cos\frac{\beta}{2} + i\sin\frac{\beta}{2}\cos\theta \end{bmatrix} = e^{i\frac{\beta}{2}}e^{-i\frac{\beta}{2}\hat{n}\cdot\vec{\sigma}}, \tag{8}$$

where $\hat{n} = (\sin\theta\cos\phi, \sin\theta\sin\phi, \cos\theta)$ is a unit vector, and $\vec{\sigma} = (\sigma_x, \sigma_y, \sigma_z)$ denotes the Pauli matrix. Eq. (8) represents a rotation around the axis $\hat{n}$ by an angle of $\beta$, up to a global phase factor $\exp(i\beta/2)$. Therefore, by selecting appropriate parameters, $\theta$, $\phi$ and $\beta$, one can construct arbitrary single qubit gates.

In theory, gate operations can be achieved by employing any type of pulses, as long as they satisfy the criteria outlined in Eq. (4) and Eq. (5). It is important to note that different pulses may exhibit significant variations in operational fidelity, as only those pulses that are robust against the system's inherent constraints or limitations can achieve a high level of operational fidelity [30,32,34]. Consequently, the ansatz of the pulses plays a crucial role in gate operations. In the next subsection, we will propose a pulse model with multiple degrees of freedom, which will be leveraged to optimize gate performance across various physical systems.

*2.2 Ansatz of Pulses*

Based on the theoretical scheme presented above, we propose an ansatz for $\Omega(t)$ as follows,

$$\Omega(t) = \frac{0.5\pi}{\tau} + \sum_{n=1}^{\infty} \alpha_n \frac{n\pi}{\tau}\cos\left(\frac{n\pi t}{\tau}\right), \tag{9}$$

where $\alpha_n$ denotes the weights assigned to each Harmonic term. Eq. (9) automatically fulfills Eqs. (4) and (5). It is worth noting that the weights $\alpha_n$ don't have any effects on the pulse area, neither on the overall scheme of the quantum gates. But they provide us with multiple degrees of freedom that can be employed to optimize the performance of quantum gates in diverse quantum systems.

From a practical view, it is preferable for the pulses to initiate and terminate at zero at the starting and end times, i.e. $\Omega(0) = \Omega(\tau) = \Omega(2\tau) = 0$. The purpose is to avoid any abrupt change of the amplitude in time domain, so to prevent any unwanted off-resonant excitations by the unnecessary frequency components. These conditions necessitate that $\alpha_n$ fulfils the following criteria:

$$\alpha_1 + 3\alpha_3 + 5\alpha_5 + \cdots + (2k-1) \cdot \alpha_{2k-1} = 0, \tag{10}$$

$$2\alpha_2 + 4\alpha_4 + 6\alpha_6 + \cdots + (2k) \cdot \alpha_{2k} = -0.5 \ (k = 1,2,3,\ldots). \tag{11}$$

The maximum value of $k$ in Eqs. (10) and (11) are theoretically unlimited. However, in experiments the generation of light or microwave pulses typically requires the use of an arbitrary waveform generator that has limited temporal and vertical resolution. This practical constraint limits the upper frequency within the Harmonic terms. In this work, we concerned the maximum value of $k$ as 2 for an illustration.

## 2.3 Optimization of the Pulses

The performance of the pulses, particularly, the determination of the values of the multiple degrees of freedom, $\alpha_n$ in Eq. (9) is evaluated by the operational fidelity. This is achieved by numerically solving the master equation in the Lindblad form [37]. It reads as follows

$$\dot{\rho} = -i[\text{H}(t), \rho] + \frac{1}{2} \sum_{i=1,2,3} \Gamma_i L(\sigma_i), \tag{12}$$

where $\text{H}(t)$ represents the Hamiltonian involving the frequency detuning $\Delta$ and reads

$$\text{H}(t) = \frac{1}{2} \begin{bmatrix} 0 & \Omega_0(t) & 0 \\ \Omega_0^*(t) & 2\Delta & \Omega_1^*(t) \\ 0 & \Omega_1(t) & 0 \end{bmatrix}, \tag{13}$$

and $\rho$ denotes the density matrix of the systems under consideration. $\Gamma_1$ and $\Gamma_2$ represent the dephasing rates between $|e\rangle$ and the qubit levels, respectively. $\Gamma_3$ denotes the dephasing rate between the two qubit levels. In this work we set $\Gamma_3 = 0$ because the coherence time of $|0\rangle \leftrightarrow |1\rangle$ transition is significantly longer than the coherence times of transitions $|0\rangle \leftrightarrow |e\rangle$ and $|1\rangle \leftrightarrow |e\rangle$ [27,39]. These shorter coherence times ultimately limit the gate operational fidelity, making the impact of $\Gamma_3$ negligible in our analysis. $L(\sigma_i) = 2\sigma_i \rho \sigma_i^\dagger - \sigma_i^\dagger \sigma_i \rho - \rho \sigma_i^\dagger \sigma_i$ denotes the Lindblad operator, where $\sigma_1 = |0\rangle\langle e| + |1\rangle\langle e|$, $\sigma_3 = |0\rangle\langle 1|$ and $\sigma_2$ varies depending on the specific level structure. Specifically, $\sigma_2 = |e\rangle\langle e| - |0\rangle\langle 0| - |1\rangle\langle 1|$ in the REI system ($\Lambda$ system) [40], and $\sigma_2 = 2|e\rangle\langle e| - |0\rangle\langle 0| - |1\rangle\langle 1|$ in the superconducting qubit system [41].

In this work, we suppose the qubit is initially in state $|\psi_{in}\rangle = |1\rangle$ for simplicity, then take the NOT gate with $(\theta, \phi, \beta) = (\pi/2, 0, \pi)$ and Hadamard gate with $(\theta, \phi, \beta) = (\pi/4, 0, \pi)$ as examples to evaluate the performance of quantum gates. The operational fidelity is defined as $F = \langle \psi_{tar}|\rho|\psi_{tar}\rangle$, where $|\psi_{tar}\rangle$ denotes the target state. $|\psi_{tar}\rangle = U(4\tau, 0)|\psi_{in}\rangle$ for ensemble REI and superconducting circuit systems and $|\psi_{tar}\rangle = U(2\tau, 0)|\psi_{in}\rangle$ for single REI system. The genetic algorithm (GA) was employed to search for the optimal values of $\alpha_n$ due to its ability to handle multi-objective functions and avoid local minima. GA is a widely used computational model that emulates natural selection and Darwinian evolutionary mechanisms, making it effective for exploring large, complex search space. For comparison purposes, we also tested a random set of parameters $\alpha_n$ to examine the effectiveness of the GA. Further details on the optimization process are provided in Appendix A.

In the subsequent section, we wifll show the numerical simulation results of the optimized gate operations in response to the different constraints in three distinct physical systems.

## 3. Simulation Results

In this section, we will present the numerical simulation results of the gate operations by employing the theoretical scheme described in Section 2 to three typical experimental quantum systems: the ensemble REI qubits system, the single REI qubit system, and the superconducting transmon qubit system in response to different inherent constraints.

### 3.1 Ensemble REI Qubits

As described in Introduction, high-fidelity gate operations in ensemble REI system requires two conditions. First, the gate operations must demonstrate resilience to frequency detuning within an interval of 170 kHz around the center frequency. To achieve this robustness, a second pair of pulses, called compensation pulses, should be implemented within the time period $[2\tau, 4\tau]$. The need for compensation arises because, as the bright state $|b\rangle$ of the ensemble ions evolves along the paths on the Bloch sphere (as shown in Fig. 2), it accumulates not only a geometric phase $e^{i\beta}$ but also an additional detuning-dependent dynamical phase. The compensation pulses are specifically designed to drive the dark state $|d\rangle$ up to $|e\rangle$ and then back down to $|d\rangle$ along the same path on the Bloch sphere. This ensures that $|d\rangle$ attains the same dynamical phase as the $|b\rangle$ state [42]. By aligning the phase accumulation for both states, we can treat this dynamical phase as a global phase factor. This compensation enhances the robustness of the gate operational fidelity against frequency detuning. Further details regarding the implementation of the compensation pulses are provided in Appendix B. Second, the gate operations should keep the unwanted off-resonant excitations at a reasonably low level. Achieving both high robustness and low off-resonant excitations simultaneously is challenging due to their contradictory nature. Therefore, a trade-off is necessary between achieving high robustness and minimizing off-resonant excitations. This trade-off is managed by optimizing the values of $\alpha_n$ in Eq. (9) for a NOT gate. The optimal values are presented in Table 1. For the optimization $\tau = 0.75$ μs, $\Gamma_1 = 2\pi \times 0.97$ kHz and $\Gamma_2 = 2\pi \times 1.21$ kHz, corresponding to $T_1 = 164$ μs and $T_2 = 132$ μs [29], respectively.

Table 1. Optimal values obtained in a NOT gate for three different systems.

| Gate | Experimental systems | $\alpha_1$ | $\alpha_2$ | $\alpha_3$ | $\alpha_4$ |
|---|---|---|---|---|---|
| NOT gate | Ensemble REI qubits | -0.6955 | -0.1966 | 0.2318 | -0.0267 |
|  | Single REI qubits | -0.0096 | -0.1317 | 0.0032 | -0.0586 |
|  | Superconducting transmon qubits | -0.8000 | -0.0365 | 0.2667 | -0.1068 |

With the optimal values of $\alpha_n$ and $\Delta=$ 170 kHz, the time evolution of Rabi frequency (see Fig. 3(a) and 3(c)) and the population of the qubit state (see Fig. 3(b) and 3(d)) for a NOT and a Hadamard gate are shown in Fig. 3. The dotted lines in (3b) and (3d) denote the evolution of the operational fidelity $F$. All the populations at the final time of the pulses align with expectations. While the $\alpha_n$ parameters were initially optimized for the NOT gate, we found that they performed effectively well for the Hadamard, $\sigma_y$ and $\sigma_z$ gates (see Table 2 in Appendix C). We have also performed individual optimizations for each gate, and found that the fidelity varies by no more than 0.15% (see Table 3 in Appendix C).

The dependence of the operational fidelity $F$ and the off-resonant excitation $P(|\Delta| \geq 3.5$ MHz) on frequency detuning $\Delta$ was investigated to evaluate the gate performance. The results are shown in Fig. 4(a) and 4(b) for the NOT gate, and Fig. 4(c) and 4(d) for the Hadamard gate. The average fidelity within the frequency detuning range of ±300 kHz is 98.09% for the NOT gate, and 98.38% for the Hadamard gate. In comparison, the average fidelity attained with randomly selected parameters, $\alpha_1 \sim \alpha_4 = (0, -0.25, 0, 0)$ within the same frequency detuning range is only 95.88% for the NOT gate and 98.01% for the Hadamard gate.

The off-resonant excitations, characterized by the population in levels $|0\rangle$ and $|e\rangle$ at $|\Delta| \geq 3.5$ MHz, are below 4% for the NOT gate and 5% for the Hadamard gate. In the case with the randomly selected parameters mentioned above, the off-resonant excitations are slightly lower than those in the optimal case.

In summary, the numerical simulation results show that the fidelity of the gate operations reaches up to 98.09% and 98.38% across a frequency detuning range of ±300 kHz, with off-resonant excitations (>3.5 MHz) at 4% and 5% for the NOT and Hadamard gate, respectively. This level of robustness and minimal off-resonant excitation meets the requirements for high-fidelity qubit manipulation in REI systems [30,43]. Compared to the average gate fidelity of 99% reported in [42] for a ±410 kHz frequency detuning range, where decay and dephasing were not considered, the average fidelity presented here is approximately 1% lower. The reduction is attributed to decoherence effects during the pulse operation time. When decay and dephasing are disabled in the simulation, the fidelity increases to 99.84%.

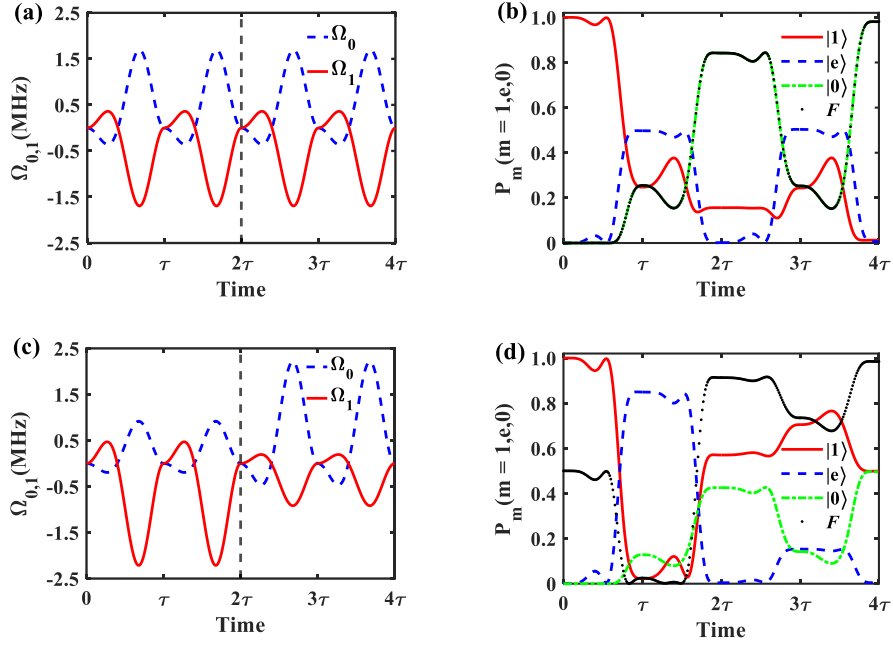

Fig. 3. Time evolution of the Rabi frequency $\Omega_{0,1}$ and the population $P_m$ for a NOT gate (a and b) and a Hadamard gate (c and d) with the optimal values of $\alpha_n$ and $\Delta = 170$ kHz in an ensemble REI system. The vertical dashed lines at $2\tau$ in (a) and (c) indicate the intersection between the qubit operation pulse and the compensation pulse. $P_m (m = 1, e, 0)$ represents the population in state $|1\rangle$, $|e\rangle$ or $|0\rangle$.

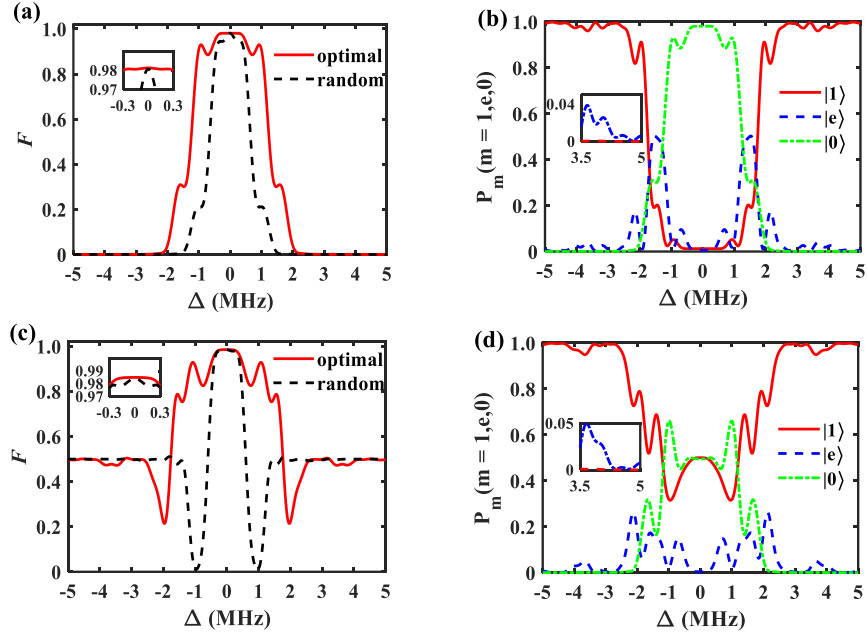

Fig. 4. Dependence of the operational fidelity $F$ and the off-resonant excitation $P (|\Delta| \geq 3.5$ MHz$)$ at $t = 4\tau$ on frequency detuning $\Delta$ for a NOT gate (a and b) and a Hadamard gate (c and d). The red-solid (black-dashed) lines

denote the result with optimal parameters (randomly selected parameter).

All the fidelities reported above are for the initial state $|1\rangle$ and may vary slightly with different initial states. To account for this variation, we calculated the average fidelity ($F_{sim}$) by simulating 2601 initial states uniformly distributed on the Bloch sphere. The simulation yielded $F_{sim} \approx 97.7\%$, which aligns closely with the theoretical prediction of $F_{cal} = 97.9\%$, bounded by the given $T_1$ and $T_2$ values in a three-level system [44]. Moreover, $F_{sim}$ can be improved to 97.8% using an alternative set of pulse parameters ($\alpha_1 \sim \alpha_4$ = -0.1547, -0.5553, 0.0516, 0.1527), which are optimized solely for fidelity at $\Delta = 0$, without imposing constraints on off-resonant excitations.

In all the simulation results reported above, four harmonic terms were used. Based on the analysis in the article [45], the number of degrees of freedom ($m$) in the control pulse should at least correspond to the dimension ($N$) of the quantum system as $m \geq 2N - 2$. For a three-level system ($N = 3$), this implies $m \geq 4$. Additionally, considering the constraints on the coefficients $\alpha_n$, given in Eqs. (10) and (11), the total number of harmonics should be six, corresponding to $\alpha_1 \sim \alpha_6$. To explore this further, we evaluated the average gate fidelity as a function of the number of harmonics used in the optimization, as shown in Fig. 5a (the explicit values of $\alpha_n$ are provided in Table 4). The fidelities fluctuate with a maximum deviation of 0.09% around 97.73%, without exhibiting a clear trend of improvement with the increasing number of harmonics. Furthermore, as the number of harmonics rises, the Rabi frequency exhibits rapid temporal variations (see Fig. 5b, where six harmonics are used). These rapid variations impose stringent requirements on the time resolution of equipment, such as arbitrary waveform generators, for accurate signal generation. Considering the practical limitations of signal generation and the lack of significant fidelity improvement with additional harmonics, we consider it reasonable to use $\alpha_1 \sim \alpha_4$.

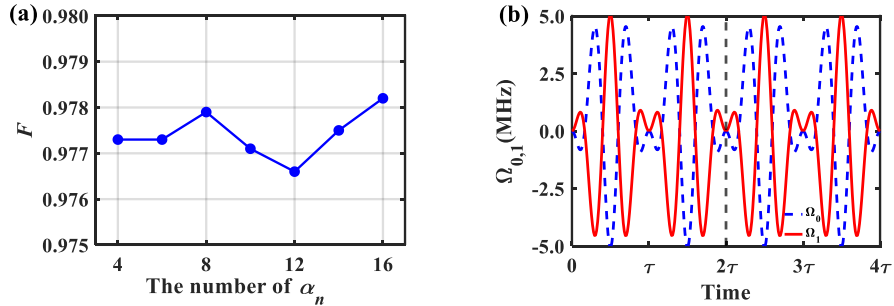

Fig. 5. Fidelity as a function of the number of harmonics (a) and the time evolution of Rabi frequency (b) where $\alpha_1 \sim \alpha_6$ = (0.0280, 0.1902, 0.0070, -0.7983, -0.0098, 0.3854) are used.

### 3.2 Single REI Qubits

The gate operation scheme presented in this work can also be applied to the single REI system,

where off-resonant excitations in the qubits, rather than the target, can be minimized through optimization of the light pulses. The optimal values of $\alpha_n$ are shown in the second row of Table 1. In the optimization process, the decay rate and dephasing rates are set to $\Gamma_1 = 2\pi \times 80$ Hz and $\Gamma_2 = 2\pi \times 60$ Hz, corresponding to $T_1 = 1.9$ ms and $T_2 = 2.6$ ms [33], with a pulse duration $\tau$ of 1 μs. The duration is slightly longer than that used in the ensemble-qubit case, as a shorter duration would result in a broader Fourier spectrum, thereby increasing off-resonant excitation.

The time evolution of Rabi frequency and population in each state is depicted in Fig. 6(a) and 6(b) for the NOT gate, and Fig. 6(c) and 6(d) for the Hadamard gate. The instantaneous Rabi frequency for both gates does not exceed 0.8 MHz, which is easily achievable in the experiments. The black-dotted lines in Fig.6(b) and Fig. 6(d) represents the evolution of the fidelity, which exceeds 99.88% by the end of the evolution.

Ideally, it is expected that the gate operations should exclusively interact with the target qubit, without affecting neighboring qubits. Thus, qubits close to the target in frequency are expected to remain in their initial state, $|1\rangle$. Hence, the sum of the unexpected population in $|0\rangle$ and $|e\rangle$ at the final time is defined as the off-resonant excitation, $P_{\text{off}} = P_0 + P_e$. The results are presented as the red-solid lines in Fig. 7(a) and 7(b). The level of off-resonant excitation at detuning $|\Delta| \geq 8.9$ MHz is found to be less than 0.5% for the NOT gate and 0.3% for the Hadamard gate, showing a decrease of about 30% comparing to instances where $\alpha_n$ was randomly selected $\alpha_1 \sim \alpha_4 = (0, -0.25, 0, 0)$, as indicated by the dashed lines.

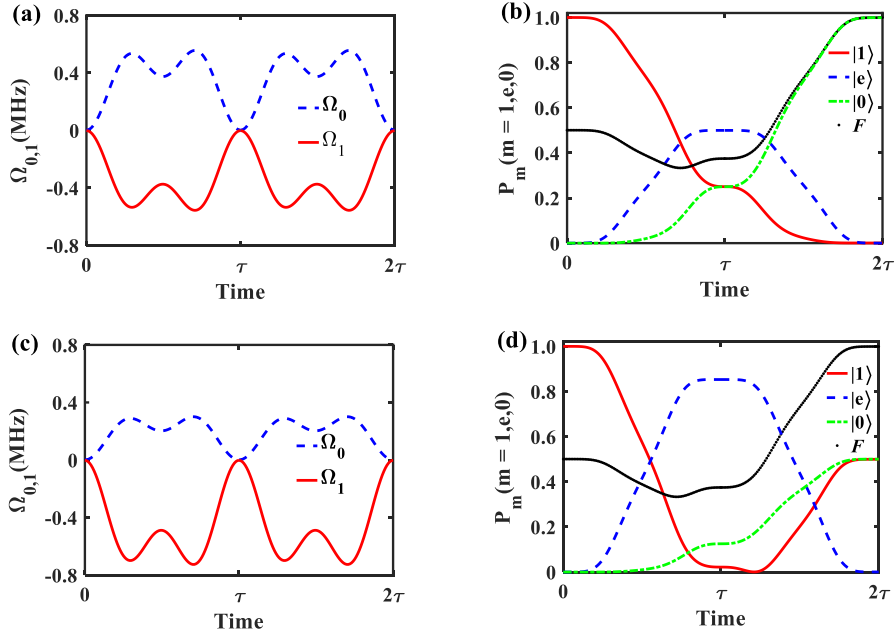

Fig. 6. Time evolution of the Rabi frequency $\Omega_{0,1}$ and the population $P_m$ for a NOT gate (a and b) and a Hadamard gate (c and d) in a single REI system.

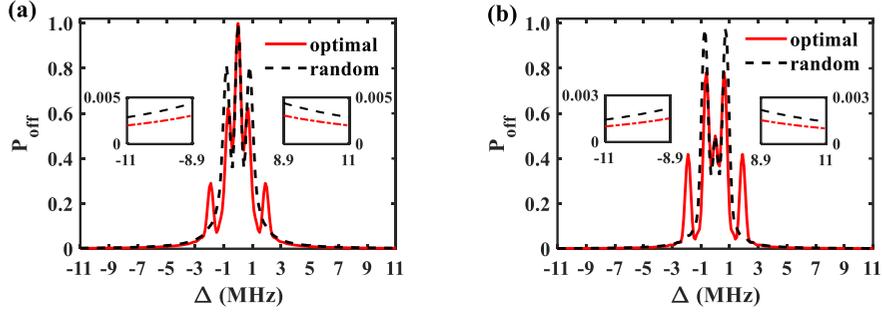

Fig. 7. Dependence of the off-resonant excitation $P_{off}$ at $t = 2\tau$ on frequency detuning $\Delta$ for a NOT gate (a) and a Hadamard gate (b).

The simulation results indicate that optimal pulses can effectively suppress off-resonant excitation beyond ±8.9 MHz, with less than 0.5% (0.3%) of the population transferred to the $|0\rangle$ and $|e\rangle$ states from $|1\rangle$ state for the NOT and Hadamard gates. Although this level of off-resonant excitation is not as low as the single-qubit gate fidelity error of approximately 0.02% reported in [33], it remains within an acceptable range. This is particularly relevant when using REI with larger energy level spacings, such as $Eu^{3+}$ in this study.

### 3.3 Superconducting transmon Qubits

A superconducting transmon qubit can also be modeled as a three-level system, as shown in Fig. 8. In contrast to the REI qubit system, microwave pulses are employed to address the qubit levels, resulting in operation time on the nanosecond scale. Here, we consider $\tau = 40$ ns with decay rates $\Gamma_1 = \Gamma_2 = 2\pi \times 3$ kHz, which are easily accessible with current technologies [34]. It is worth noting that a pair of compensation pulses must be applied in order to achieve high robustness against frequency detuning. The rationale for using compensation pulses is the same as in the ensemble REI qubits case (Section 3.1).

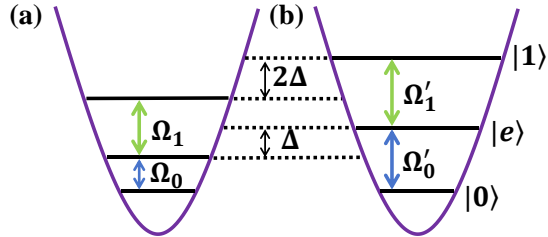

Fig. 8. Schematic representation of energy levels and Rabi frequencies for two superconducting transmon qubits, (a) and (b). The addressing frequency of qubit (b) is detuned from that of qubit (a) by $\Delta$, resulting from variations in the fabrication process of the superconducting circuits.

A set of optimal values for $\alpha_n$ is presented in the third row of Table 1 for the superconducting transmon qubits system. The time evolution of Rabi frequency, population and operational fidelity is shown in Fig. 9(a) and 9(b) for the NOT gate, and Fig. 9(c) and 9(d) for the Hadamard gate with detuning $\Delta = 2$ MHz. The Rabi frequency exhibits gradual temporal variation, and by the end of operation, the population distribution of the quantum system aligns perfectly with the desired outcome for the quantum gate.

To access the robustness of the quantum gates, we analyzed the dependence of operational fidelity $F$ on the frequency detuning $\Delta$. The results are depicted by the red-solid curves in Fig. 10(a) and 10(b) for the NOT and Hadamard gates, respectively. The black-dashed curve represents the case with randomly selected parameters for comparison. Within a frequency range of ±9.3 MHz for the NOT gate and ±12.7 MHz for the Hadamard gate, the fidelity with optimal pulses exceeds 99.6%. This range corresponds to approximately 8% of the maximum Rabi frequency and highlights the gate's robustness to frequency variations. For both gates, the robustness with optimal parameters is significantly greater than with randomly selected parameters.

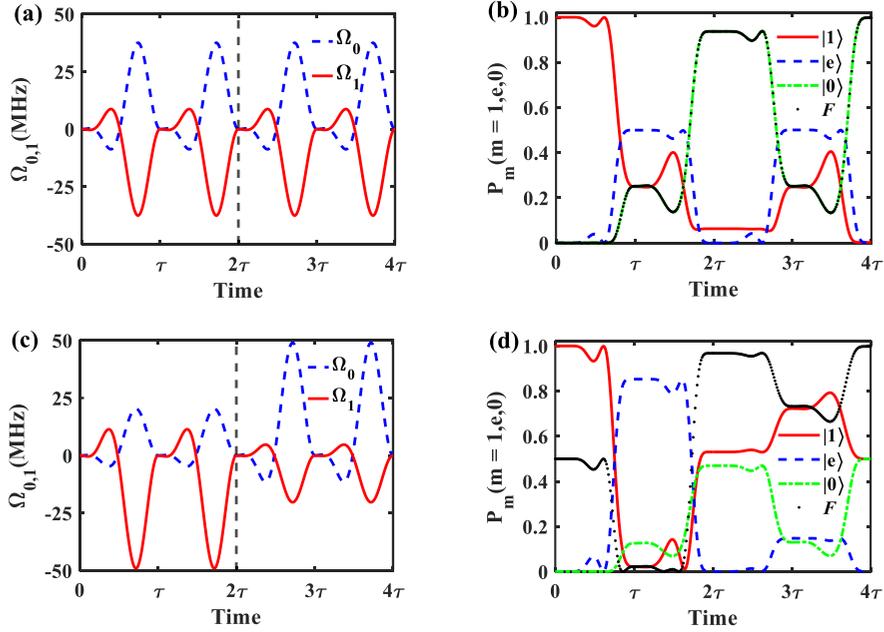

Fig. 9. Time evolution of the Rabi frequency $\Omega_{0,1}$ and the population $P_m$ for a NOT gate (a and b) and a Hadamard gate (c and d) with the optimal values of $\alpha_n$ and $\Delta = 2$ MHz in a superconducting transmon qubit system.

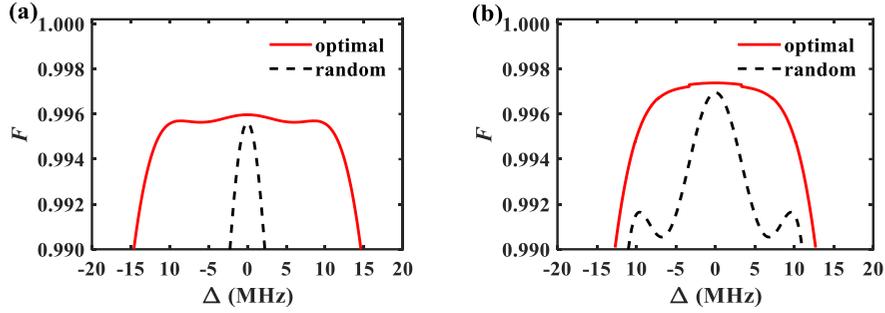

Fig. 10. Dependence of gate fidelity at $t = 4\tau$ on frequency detuning; (a) the NOT gate, (b) the Hadamard gate.

We also investigated how the infidelity (1-$F$) of the NOT gate and Hadamard gate is affected by variations in the optimal parameters ($\eta\alpha_{1\sim4}, |\eta| \leq 0.3$) at different frequency detuning levels ($|\Delta| \leq \pm2$ MHz). The results are shown in Fig. 11(a) through 11(d) for variations in $\alpha_1$, $\alpha_2$, $\alpha_3$ and $\alpha_4$, respectively. For $\alpha_1$, a ±30% variation results in an increase in infidelity of 0.7% for the NOT gate, and 0.17% for the Hadamard gate at $\Delta = \pm2$ MHz. However, variations of ±30% in $\alpha_2$, $\alpha_3$ and $\alpha_4$ lead to only a 0.1% increase in infidelity for the NOT gate and 0.01% for the Hadamard gate. This is due to the significant contribution of the first Harmonic term, which is weighted by $\alpha_1$, in pulse construction. If $\alpha_1$ experience a variation of ±8% or ±13%, the same increase in infidelity for the NOT and Hadamard gate as a ±30% in $\alpha_2$, $\alpha_3$, and $\alpha_4$.

In summary, our simulation results demonstrate that both the NOT and Hadamard gates exhibit strong robustness against variations in pulse intensity and frequency detuning. Gate operation fidelity reaches up to 99.6% within a frequency detuning range of ±9.3 MHz for the NOT gate and ±12.7 MHz for the Hadamard gate, showing a noticeable improvement over the previously reported 98% fidelity within a ±1.5 MHz range in [46]. Furthermore, even with up to a 30% variation in the four pulse parameters, the infidelity increases marginally, with a maximum rise of 0.7% for the NOT gate and 0.1% for the Hadamard gate. These results confirm that the gates maintain high fidelity under experimental conditions, demonstrating their reliability for quantum computing applications.

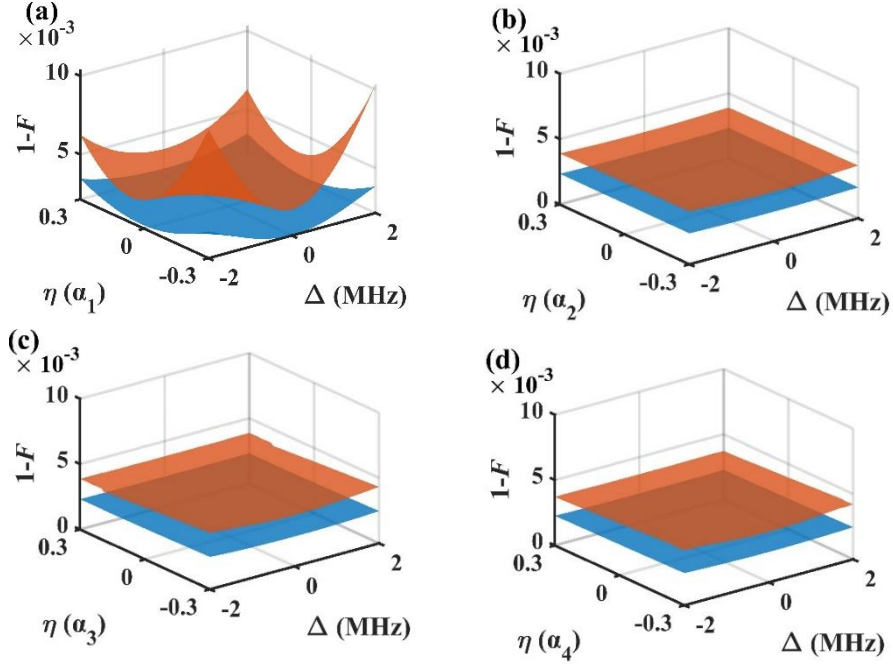

Fig. 11. The infidelity of the NOT (orange curved surface) and the Hadamard gate (blue curved surface) with the fluctuation $\eta$ in $\alpha_1 \sim \alpha_4$ and frequency detuning $\Delta$.

## 4. Discussion and Conclusion

In this work, we have proposed a versatile theoretical scheme that utilizes multiple degrees of freedom to implement universal NHQC gate operations.

We applied this scheme to three commonly used qubit systems: ensemble REI qubits, single REI qubits, and superconducting transmon qubits system. For the ensemble REI qubit system, the optimized gate operations achieve robust fidelity (greater than 98%) against frequency detuning within ±300 kHz and induce low off-resonant excitation (<5%) beyond ±3.5 MHz. In the single REI system, gate fidelity reaches up to 99.88% with negligible off-resonant excitations (<0.5%). In the superconducting transmon qubit system, operational fidelity can reach 99.6% within a frequency detuning range of ±9.7 MHz. This robustness could significantly reduce the workload in the surface code quantum computing approach.

Currently, all gates in this approach have the same time costs, regardless of the geometric phases associated with each gate. Future work will address this by incorporating phase-dependent durations in the approach. Additionally, this approach can be extended to two-qubit gates, such as the CNOT, by enabling precise, independent control over each qubit's state.

In summary, our adaptable scheme demonstrates potential for advancing high-fidelity quantum gates across systems, offering a robust foundation for practical quantum computing platforms.

**APPENDIX A: OPTIMIZATION OF PULSE PARAMETERS**

Here, we take the ensemble REI system as an example to demonstrate how GA was applied to optimize the multiple degrees of freedom in the light pulse. GA is an optimization technique based on the principles of natural selection and genetics. GA works by simulating the process of evolution through generations of potential solutions, improving solutions using mechanisms: selection, crossover, and mutation.

For our optimization, the pulse parameters $\alpha_1 \sim \alpha_4$ (as shown in Eq. 9) are encoded as chromosomes. The initial population consists of 50 randomly generated chromosomes within the target range [-0.8, 0.8]. The fitness of each generation is determined using a combination of Pareto dominance and crowding distance. Higher fitness chromosomes are selected and randomly paired for crossover to generate the next generation, with evolution continuing for 300 generations.

The optimization minimizes two objective functions: one targeting a low average infidelity (1-$F$) within the range from -170 kHz to 170 kHz, and the other targeting a low average off-resonant excitations, characterized by the overall population in the $|0\rangle$ and $|e\rangle$ states within the frequency detuning range of 3.5 MHz to 5 MHz. These objectives are evaluated concurrently. However, they are inherently conflicting, *i.e.* suppressing one leads to an increase in the other, as both objectives are affected similarly by frequency detuning.

As a result, there is no single fitness value that summarizes the overall performance. Instead, the fitness function uses a dual ranking mechanism based on Pareto dominance and crowding distance to explore diverse Pareto-optimal solutions. The top 30% of these solutions form the Pareto front, see Fig. 12, which represents the best trade-offs between conflicting objectives.

Each point on the Pareto corresponds to a chromosome (one set of parameters $\alpha_1 \sim \alpha_4$). The sixth point from the left on the Pareto front is selected as the optimal solution for the ensemble REI system, with $\alpha_1 \sim \alpha_4$ listed in the first row of Table 1.

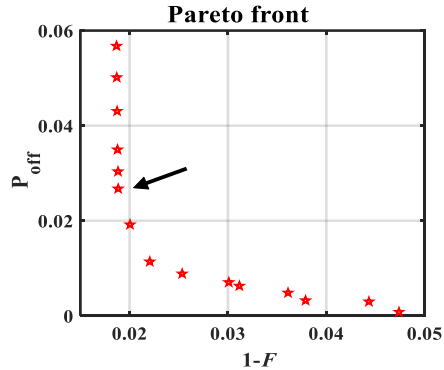

Fig. 12. Pareto front of the optimization results in the ensemble REI system. The sixth point from left to right, indicated by the arrow, denotes the optimal solution. 1-$F$ represents the average infidelity over the detuning range of -170 kHz to 170 kHz, while $P_{off}$ represents the average excitation in the range of 3.5 MHz to 5 MHz.

The entire optimization process consists of five sequential steps as follows:

(i) Define the initial state: start with $|\psi_{in}\rangle = \cos\theta\,|0\rangle + \sin\theta\,e^{i\varphi}|1\rangle$. In this work, we use $|1\rangle$ as the initial state for simplicity.

(ii) Define the pulses: specify the pulses $\Omega_{0,1}$ based on parameters $\alpha_n$.

(iii) Extract the density matrix: solve the Lindblad master equation to obtain the density matrix $\rho$.

(iv) Calculate the operational fidelity at the final time: $F=|\langle\psi_{tar}|\rho|\psi_{tar}\rangle|^2$, where $|\psi_{tar}\rangle = U(4\tau,0)|\psi_{in}\rangle$ represents the desired target state of the gate operations.

(v) Optimize the pulse parameters: use the GA as described above to optimize the pulse parameters until they meet the system requirements.

We believe GA is the more efficient and robust choice for the multi-objective optimization tasks in this work. While alternative approaches, such as the goal attainment method, could yield similar results if provided with appropriate initial values, GA is more adaptable and reliable. It requires less dependency on parameter selection, and is capable of exploring the solution space more comprehensively, ultimately finding high-quality solutions with reduced computational overhead.

**APPENDIX B: COMPENSATION PULSES**

A pair of compensation pulses are used to enhance the robustness of quantum manipulation for ensemble REI qubits and superconducting transmon qubits. The light/microwave drives the states evolving from $|d\rangle$ to $|e\rangle$ during the time period of $t \in [0,2\tau]$, then back to $|d\rangle$ along the same path during the time period $t \in [2\tau, 4\tau]$. The change in interaction is achieved by transforming the original $|d\rangle$ to a new bright state by setting the parameters $(\theta', \varphi_0', \varphi_1')$ as follows:

$$\theta' = \pi - \theta$$
$$\varphi_0' = \begin{cases} -(\pi+\phi), & t \in [2\tau, 3\tau] \\ -\phi, & t \in [3\tau, 4\tau] \end{cases} \quad (14)$$
$$\varphi_1' = \begin{cases} 0, & t \in [2\tau, 3\tau] \\ \pi, & t \in [3\tau, 4\tau] \end{cases}$$

In addition to Eq. (14), the compensation pulse must also satisfy the pulse area condition shown in Eqs. (4) and (5). We propose the Rabi frequency as follows:

$$\Omega'(t) = \frac{0.5\pi}{t} + \sum_{n=1}^{\infty} \alpha_n' \frac{n\pi}{\tau} \cos\left(\frac{n\pi t}{\tau}\right) \quad (15)$$

where $\alpha_n'$ should meet the following boundary conditions, similar to $\alpha_n$ in Eqs. (10) and (11):

$$\alpha_1' + 3\alpha_3' + 5\alpha_5' + \cdots + (2k-1)\cdot\alpha_{2k-1}' = 0, \quad (16)$$
$$\alpha_2' + 2\alpha_4' + 3\alpha_6' + \cdots + (2k)\cdot\alpha_{2k}' = -0.25\ (k=1,2,3,\dots). \quad (17)$$

In theory, $k$ in Eqs. (16) and (17) can be infinitely large, and $\alpha_n'$ can differ from $\alpha_n$. For simplicity, we set $\alpha_n' = \alpha_n$, and used the pulses in Eqs. (9)-(11) and Eqs. (15)-(17) in the numerical simulations described in Section 3.

**APPENDIX C: PERFORMANCE OF OTHER GATES**

In this appendix, we first present the performance of $\sigma_y$ and $\sigma_z$ gates in the three systems using the same pulse parameters as shown in Table 1, which were optimized for the NOT gate. For both gates, the fidelity consistently exceeds 98% over a frequency detuning, while the off-resonant excitation remains below 5%, as shown in Table 2. For comparison, data for the NOT and Hadamard gates are also provided.

**Table 2. Performance metrics of quantum gates for different systems using the parameters optimized for a NOT gate.**

| Gates | Ensemble REI qubits | | Single REI qubits | Superconducting qubits |
|---|---|---|---|---|
| | Fidelity | Off-resonant excitation | Off-resonant excitation | Fidelity |
| NOT | 98.09% | 4% | 0.5% | 99.76% |
| Hadamard | 98.38% | 5% | 0.3% | 99.79% |
| $\sigma_y$ | 98.06% | 4% | 0.5% | 99.76% |
| $\sigma_z$ | 98.85% | 5% | 0.1% | 99.85% |

Beyond the simulation results presented in Table 2, we also conducted individual optimizations for each quantum gate, with the results summarized in Table 3. The fidelity varies by no more than 0.15%, and off-resonant excitation varies by 1% when compared to the performance using the optimal parameters obtained from a NOT gate. These results validate the robustness and effectiveness of our optimization scheme across different types of quantum gates.

**Table 3. Performance metrices of quantum gates for the ensemble REI system using parameters optimized for each individual gate.**

| Gates | Optimized parameters $\alpha_1 \sim \alpha_4$ | Ensemble REI qubits | |
|---|---|---|---|
| | | Fidelity | Off-resonant excitation |
| NOT | -0.6955, -0.1966, 0.2318, -0.0267 | 98.09% | 4% |
| Hadamard | -0.7338, 0.0024, 0.2449, -0.1261 | 98.50% | 4% |
| $\sigma_y$ | -0.8000, -0.0753, 0.2667, -0.0873 | 98.12% | 4% |
| $\sigma_z$ | 0.7261, -0.0631, -0.2420, -0.0934 | 98.69% | 4% |

Table 4. The values of $\alpha_n$ used for the simulation in Fig. 5a

| The number of $\alpha_n$ | Optimized parameters | | | |
|---|---|---|---|---|
| 6  | 0.0280  | 0.1902  | 0.0070  | -0.7983 |
|    | -0.0098 | 0.3854  |         |         |
| 8  | 0.8000  | -0.0133 | -0.0667 | -0.0843 |
|    | -0.1021 | -0.0092 | -0.0128 | -0.0102 |
| 10 | 0.0417  | -0.0958 | -0.0051 | -0.1655 |
|    | 0.0270  | -0.2552 | 0.0013  | 0.8000  |
|    | -0.0190 | -0.4515 |         |         |
| 12 | -0.0156 | -0.0892 | 0.0133  | -0.1770 |
|    | 0.0107  | 0.7538  | -0.0042 | -0.4066 |
|    | -0.0206 | 0.6352  | 0.0124  | -0.6029 |
| 14 | -0.0080 | -0.2069 | 0.0046  | -0.0650 |
|    | 0.0127  | -0.0687 | 0.0614  | -0.0907 |
|    | -0.0256 | -0.1017 | 0.0226  | 0.5747  |
|    | -0.0398 | -0.3262 |         |         |
| 16 | -0.8000 | -0.0000 | 0.2702  | 0.0000  |
|    | -0.0001 | 0.0000  | -0.0001 | -0.0000 |
|    | -0.0001 | -0.0000 | -0.0002 | -0.0000 |
|    | -0.0002 | 0.0000  | -0.0002 | -0.0312 |

**Acknowledgment.** The work was supported by self-determined Research Project of the Key Lab of Advanced Optical Manufacturing Technologies of Jiangsu Province (ZZ2109). Joel M. acknowledges the National Natural Science Foundation of China with grant number 62074107.

**Disclosures.** The authors declare no conflicts of interest.

**Data availability.** All data needed to evaluate the conclusions in the article are presented in the article. Additional data related to this paper may be requested from the corresponding authors.